\documentclass{anstrans}
\title{Development of a High Intensity Neutron Source at the European Spallation Source:  the HighNESS Project}

\author{V. Santoro$^{*, \ddag}$, K.H. Andersen$^{a}$, M. Bernasconi$^{b}$, M. Bertelsen$^{*}$, Y. Be{\ss}ler$^{c}$, D. Campi$^{b}$, 
V. Czamler$^{d}$, D. D. Di Julio$^{*}$, E. Dian$^{e,f}$, K. Dunne$^{g}$, P. Fierlinger$^{h}$, A. Gaye$^{*}$, G. Gorini$^{b}$, C. Happe$^{c}$, T. Kittelmann$^{*}$, E.B. Klinkby$^{i}$, Z. Kokai$^{*}$, R. Kolevatov$^{j}$, B. Lauritzen$^{i}$, R. Linander $^{*}$, J.I. Marquez Damian$^{*}$, B. Meirose$^{g,k}$, F. Mezei$^{e}$, D. Milstead$^{g}$, G. Muhrer$^{*}$, K. Ramic$^{*}$, B. Rataj$^{*}$, N. Rizzi$^{i}$, S. Samothrakitis$^{l}$, J. R. Selknaes$^{*}$, S. Silverstein$^{g}$, M. Strobl$^{l}$, M. Strothmann$^{c}$, A. Takibayev$^{*}$, R. Wagner$^{d}$, P. Willendrup$^{i}$, S.-C. Yiu$^{g}$, L. Zanini$^{*}$,  and O. Zimmer$^{d}$}

\institute{
$^{*}$ European Spallation Source ERIC, Lund Sweden
\and
$^{a}$Oak Ridge National Laboratory, Oak Ridge, TN 37831, USA
\and
$^{b}$University of Milano-Bicocca, Milano, Italy
\and
$^{c}$Forschungszentrum-Juelich, Germany
\and
$^{d}$Institut Laue-Langevin ILL, Grenoble, France
\and

$^{e}$Mirrotron Ltd., 29-33 Konkoly Thege Miklós út, 1121 Budapest, Hungary

$^{f}$ Centre for Energy Research, 29-33 Konkoly Thege Miklós út, 1121 Budapest, Hungary

$^{g}$Stockholm University, Stockholm, Sweden
\and
$^{h}$Technical University Munich, Garching, Germany
\and
$^{i}$DTU Physics, Technical University of Denmark
\and
$^{j}$ ESS consultant, Norway
\and
$^{k}$ Lund University, Lund, Sweden
\and
$^{l}$Paul Scherrer Institut, PSI Villigen, Switzerland

$^{\ddag}$\texttt{\small Valentina.Santoro@ess.eu}
}

\usepackage{graphicx} 
\usepackage{booktabs} 
\usepackage{microtype} 
\usepackage{hyperref}

\usepackage[format=hang,font={bf}]{caption}

\usepackage{caption}
\usepackage{subcaption}

\begin{document}
\section*{Abstract}
The European Spallation Source (ESS), presently under construction in Lund, Sweden, is a multi-disciplinary international laboratory that will operate the world’s most powerful pulsed neutron source. Supported by a 3M Euro Research and Innovation Action within the EU Horizon 2020 program, a design study (HighNESS) is now underway to develop a second neutron source below the spallation target. Compared to the first source, located above the spallation target and designed for high cold and thermal brightness, the new source will provide higher intensity, and a shift to longer wavelengths in the spectral regions of cold (2-20 $\AA$), very cold (VCN, 10-120 \AA), and ultra cold (UCN, > 500 \AA) neutrons.  The core of the second source will consist of a large liquid deuterium moderator to deliver a high flux of cold neutrons and to serve secondary VCN and UCN sources, for which different options are under study. The features of these new sources will boost several areas of condensed matter research and will provide unique opportunities in fundamental physics. Part of the HighNESS project is also dedicated to the development of future instruments that will make use of the new source and will complement the initial suite of instruments in construction at ESS. The HighNESS project started in October 2020. In this paper, the ongoing developments and the results obtained in the first year are described.
\bigskip

\begin{flushright}
\textbf{KEYWORDS}

cold neutrons, very cold neutrons, ultra cold neutrons, neutron instruments, fundamental physics

\end{flushright}
\section{Introduction}
The European Spallation Source (ESS) is presently under construction in Lund, once completed at full specification will be the most powerful spallation neutron source in the world. Neutrons are produced by the interaction of a 2 GeV proton beam with a tungsten spallation target. To be used by neutron instruments, the energy of the neutrons exiting the target is decreased from the MeV to the meV range, using suitable neutron moderators. ESS will include in its initial suite fifteen instruments for neutron scattering, with the accelerator intending to reach 2 MW time-average power in the initial years of operation, with a future upgrade to 5 MW. All the planned instruments, in different areas of neutron scattering like diffraction, SANS, imaging, reflectometry,
and spectroscopy, are designed to attain world-leading performance at final specification.

The first fifteen instruments will use neutrons from a moderator system placed above the spallation target, using water and liquid parahydrogen for thermal and cold neutrons, respectively. However, the ESS facility was planned with the possibility to add more instruments: thanks to a grid of 42 beamports for neutron extraction, spanning an angular range of 240$^{\circ}$, and to the presence of upgrade areas, i.e., locations kept available for the placement of additional instruments. The other key aspects for the future upgrade of ESS are the provision for the insertion of a second moderator system below the spallation target and, the beam extraction system that has been designed with the possibility to extract low-energy neutrons either from above or below the target. This configuration gives tremendous opportunities for ESS because it offers the possibility to install a second source system, with different features than the first one, and the possibility to conceive and design new and different instruments, broadening the opportunities that ESS can offer to the scientific community. The goal of HighNESS~\cite{Santoro:2020nke}, a EU-funded 3-years project started in October 2020, is to perform a design study of this second source, including a set of neutron scattering instruments and the fundamental physics experiment NNBAR~\cite{Addazi:2020nlz} (search for neutron antineutron oscillations) that will make use of the new source.
For the initial instrument suite, the focus was on a source able to deliver high brightness of thermal and cold neutrons, instead, the new sources that will be designed in the HighNESS project will focus on two different aspects: an increase of the total number of neutrons emitted from the source (source intensity), and a shift of the wavelength range towards colder neutrons. A more intense source requires larger moderators and emission surfaces, to increase the count rate for instruments or experiments that need high flux. To have a source of colder neutrons, in the HighNESS project, we foresee in addition to the second cold neutron source,  a very cold source, and an ultra cold source. This upgrade will make possible for ESS to offer to the scientific community a whole different range of neutron energies, from thermal (0.025 eV)  to ultracold neutrons (< 300 neV), for a broad range of applications and experiments.

The HighNESS project is structured in ten different Work Packages (WPs). WP1 is the {\it Project Coordination}, that is coordinating the whole activities in the project. WP2, {\it Software Development}, will develop computational tools needed to study and design the high intensity moderators. The computational tools developed by WP2 will rely on the experimental measurements performed in WP3, {\it Material Characterization with Neutrons}, to provide required
material property data needed to generate thermal-neutron scattering kernels and to validate the models. WP4 and WP6, {\it Moderators Design} and {\it Advanced Reflector}, respectively, represent the core of the neutronics design to develop the cold source and the VCN and UCN sources. These two WPs will need input and requirements from two scientific WPs (WP7,
{\it Condensed Matter Science}, and WP8, {\it Fundamental Physics}) to ensure that the designs of the sources deliver neutron fluxes and spectra appropriate to perform new and ground-breaking scientific measurements at ESS that cannot be performed with the high-brightness moderator. During this conceptual design study, there will be a continuous exchange of information between the WPs 4 and 6 devoted to neutronic design, and WP5 devoted
to {\it Engineering}, which is in charge of delivering a manufactural mechanical design. All software and data developments performed by WPs 2 and 6 will be made available to the public by a cloud computing resource developed by WP9, {\it  Computing Infrastructure}. WP10
will take care of the dissemination of all the scientific contents produced within this project.  In the next sections, the highlights from all the WPs (except for WP10 which will be covered in later publications) in the first year of the project will be described. 

\section{Work Package 2: Software Development}

The objective of WP2 is the development of simulation software for describing neutron interactions in the novel moderator and reflector materials that will be used for the design of the future ESS sources. During the first year of the HighNESS project, magnesium hydride and deuteride, along with diamond nanoparticles were analyzed since these materials are promising reflectors for cold, in the case of magnesium hydride and deuteride, and for very cold neutrons in the case of nanodiamonds. 

\begin{figure}[h]
     \centering
     \begin{subfigure}[b]{0.39\textwidth}
         \centering
         \includegraphics[width=1\textwidth]{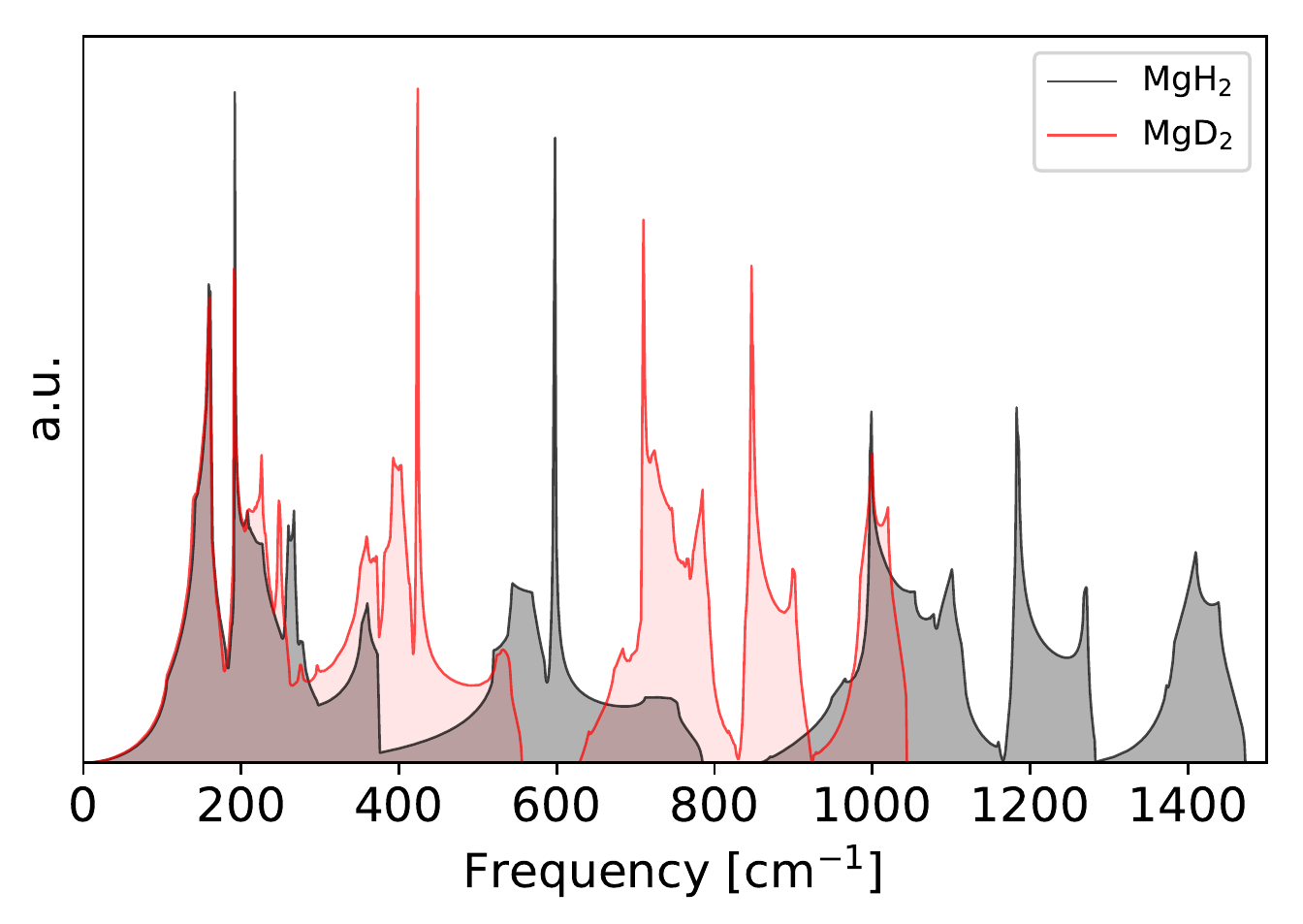} 
         \caption{}
         \label{fig:MgH2spectrum}
     \end{subfigure}
     \begin{subfigure}[b]{0.39\textwidth}
         \centering
         \includegraphics[width=1\textwidth]{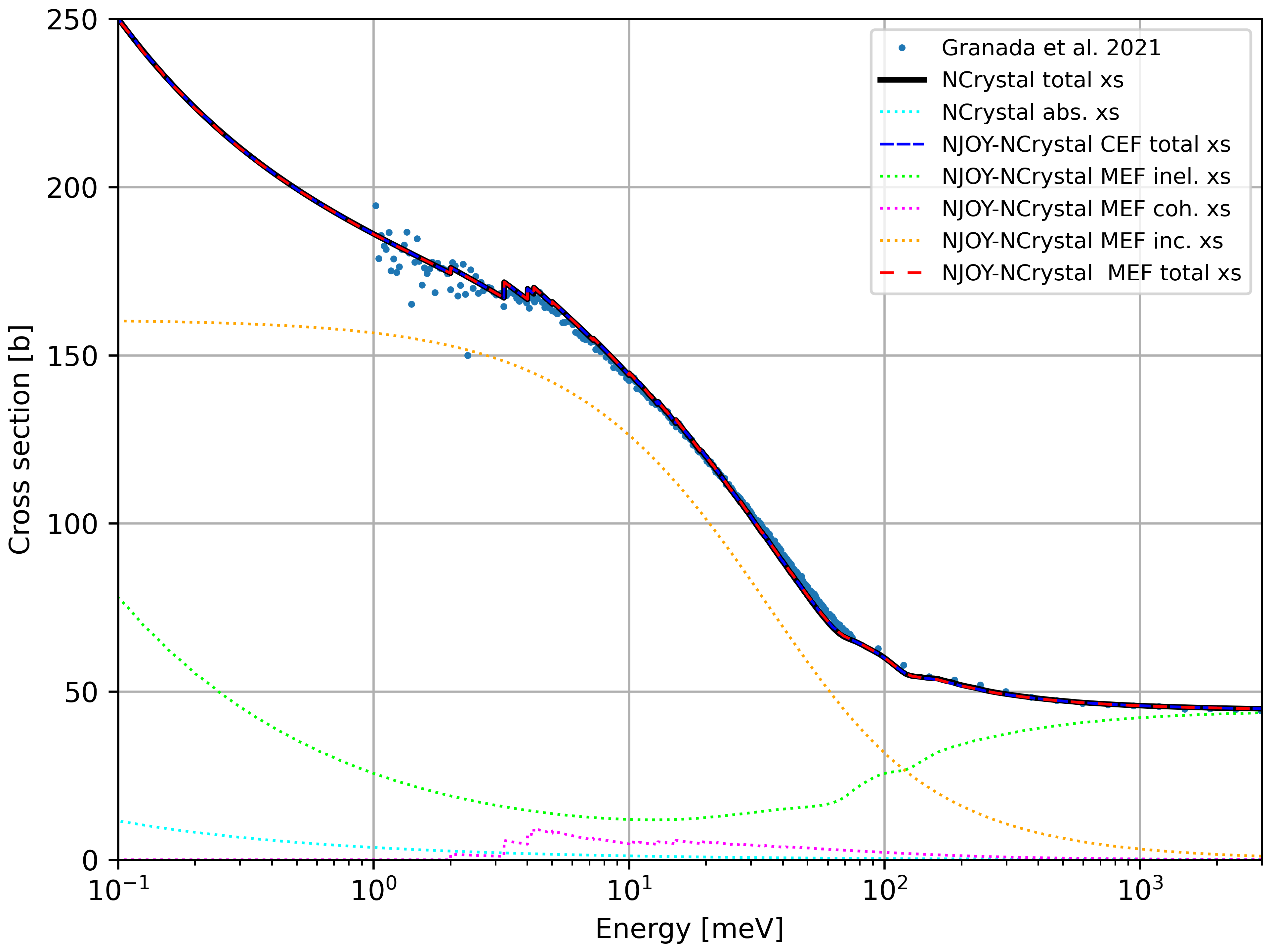} 
         \caption{}
         \label{fig:MgH2totXS}
     \end{subfigure}
        \caption{Comparison of the theoretical phonon DOS for MgH$_2$ and MgD$_2$ (left), and total cross section calculations (right) compared with experimental results from Ref. \cite{granada2021}.}
        \label{fig:MgH2}
\end{figure}

MgH$_2$ and MgD$_2$ were explored using density functional perturbation theory in order to compute the phonon spectra using the Quantum-ESPRESSO package (Fig.~\ref{fig:MgH2spectrum}). These spectra were fed into a newly developed calculation tool called NJOY+NCrystal~\cite{ncrystal}, which combines the NCrystal code developed at ESS with the standard tool NJOY to create thermal scattering libraries in ENDF-6 and ACE formats. NJOY was modified to support the newly proposed mixed elastic scattering format. Total cross section calculations for MgH$_2$ are shown in Fig.~\ref{fig:MgH2totXS}. Further details are available in a separate publication~\cite{ramic2021njoy}.

\begin{figure}[h]
     \centering
     \begin{subfigure}[b]{0.39\textwidth} 
         \centering
         \includegraphics[width=0.75\textwidth]{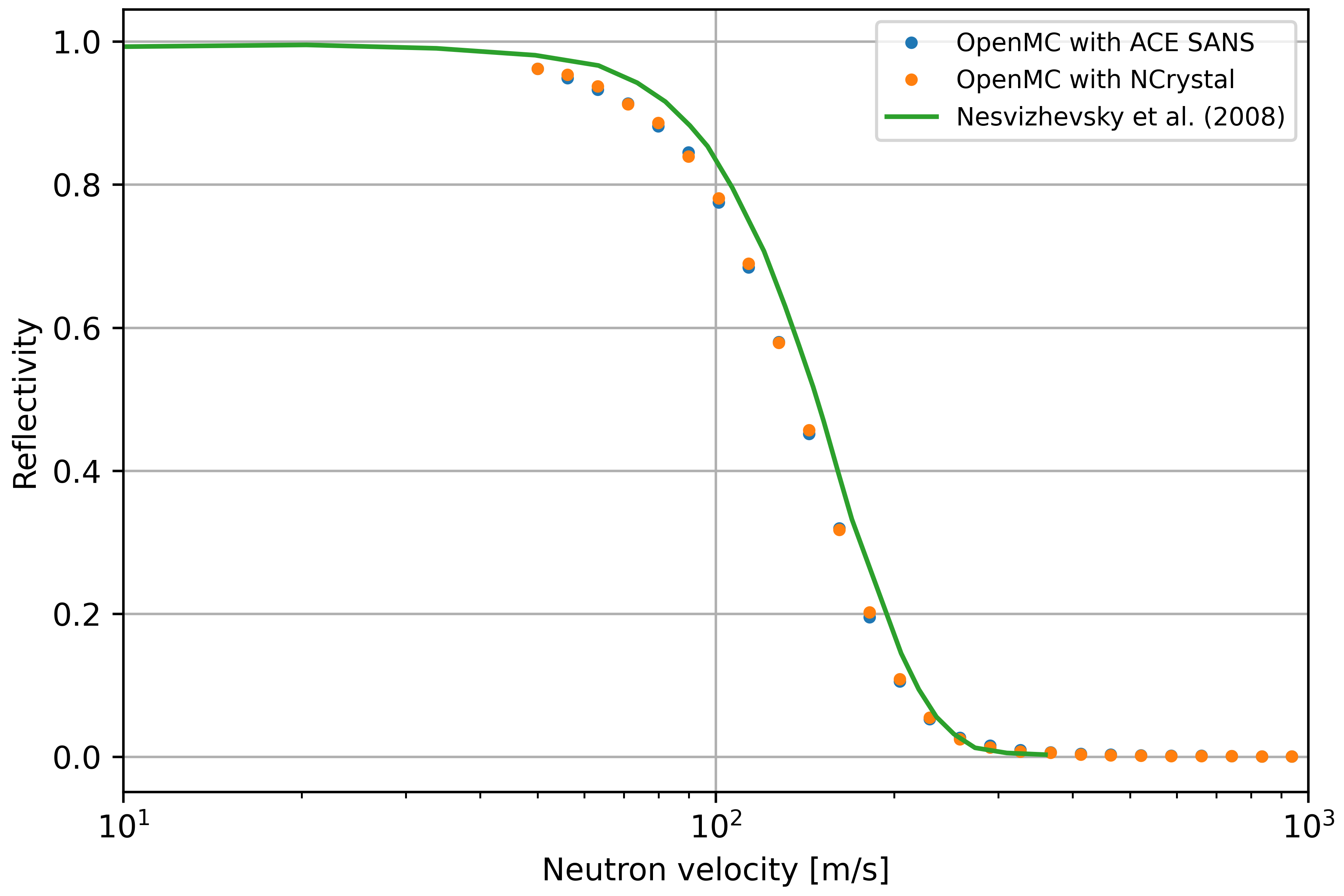}
         \caption{}
         \label{fig:NDtransmission}
     \end{subfigure}
     \begin{subfigure}[b]{0.39\textwidth} 
         \centering
         \includegraphics[width=0.95\textwidth]{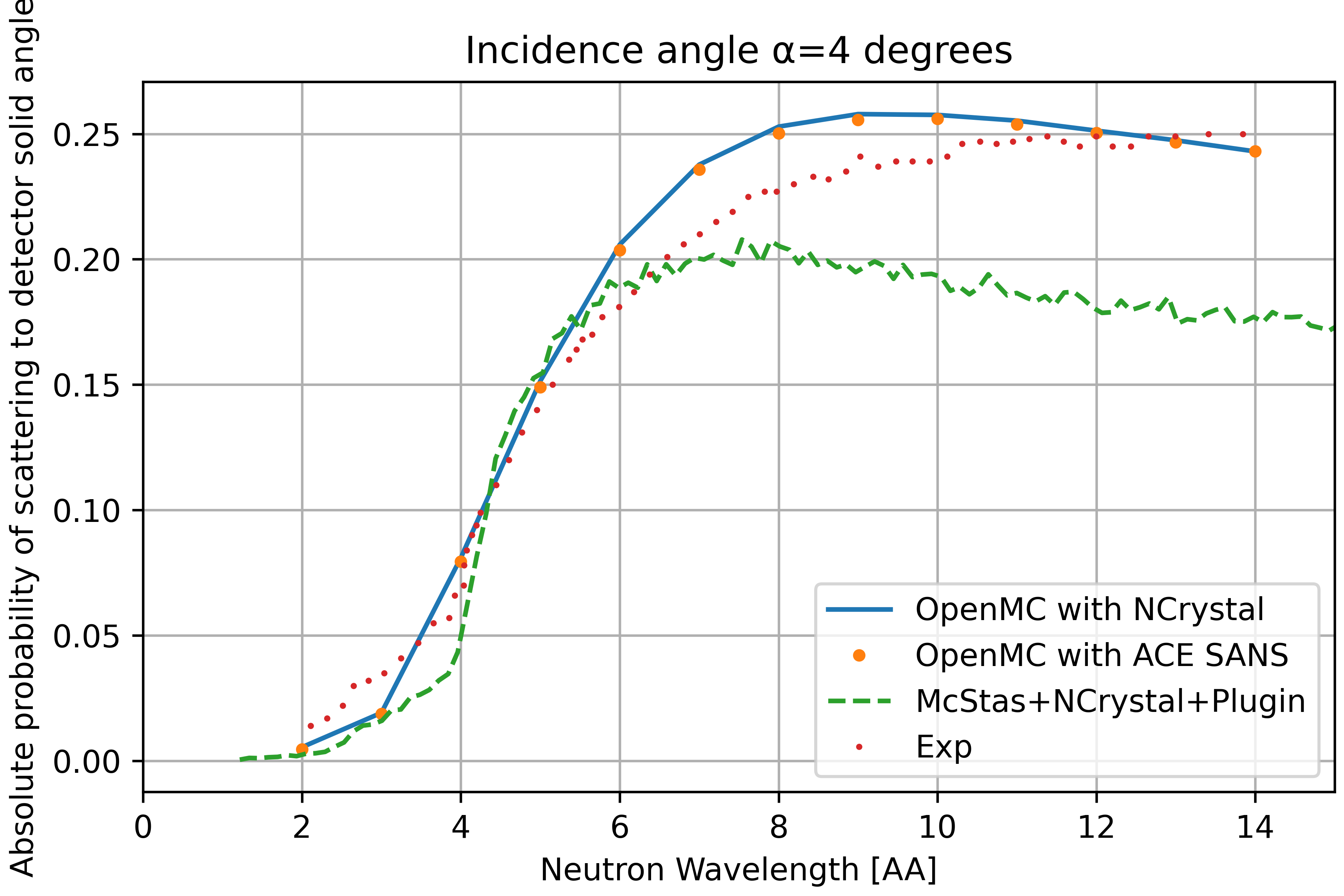}
         \caption{}
         \label{fig:NDreflectivity}
     \end{subfigure}
        \caption{Calculation of neutron total reflection probability of nanodiamonds (left) compared with results from Ref. \cite{NESVIZHEVSKY2008631}, and reflectivity (right) compared with results from Ref. \cite{cubitt_quasi-specular_2010}.}
        \label{fig:ND}
\end{figure}

For diamond nanoparticles, two lines of work were started. The team at University Milano-Bicocca started with a study of finite particle effects in nanoparticles using molecular dynamics and the AIREBO and Tersoff potentials and now continues with a study using machine learning potentials. The team at ESS, in collaboration with WP6, developed two ways to describe the small angle interaction of neutrons with nanoparticles. The first is a model implemented in the free software package NCrystal, which can be called from McStas~\cite{willendrup2004mcstas} and OpenMC, and the second is a hard coded implementation in MCNP6~\cite{mcnp}, and PHITS, which was complemented with a thermal ACE file providing Bragg and inelastic scattering. The implementations were tested using results from total reflection (Fig.~\ref{fig:NDtransmission}) and reflectivity experiments (Figs.~\ref{fig:NDreflectivity}).

All the tools generated during this work, as well as the thermal scattering libraries produced, are available in the HighNESS Github repository \cite{highnessgithub}.
\section{Work Package 3: Material Characterization with Neutrons}

The main goal of WP3 is the characterizations of materials that promise large gains in the moderation of cold neutrons down to the energy range of VCNs. To do that several experiments on various instruments of the Institut Laue-Langevin (ILL) in Grenoble, France have been and will be performed in the course of the HighNESS project.
The characterization of these materials and the subsequent use of them in neutron moderation and reflection will enhance the capabilities of numerous neutron scattering techniques as well as the scientific reach of particle-physics experiments employing beams of slow neutrons. 

Clathrate hydrates have been identified as a promising material for this purpose.  It has already been shown for a particular type, namely tetrahydrofuran (THF) clathrate, that it offers a rich spectrum of incoherent low-energy excitations~\cite{conradInelasticScatteringSpectral2004}, a property that could be utilized to moderate neutrons to lower energies.

\begin{figure}[h]
     \centering
     \begin{subfigure}[b]{0.48\textwidth}
         \centering
         \includegraphics[width=0.8\textwidth]{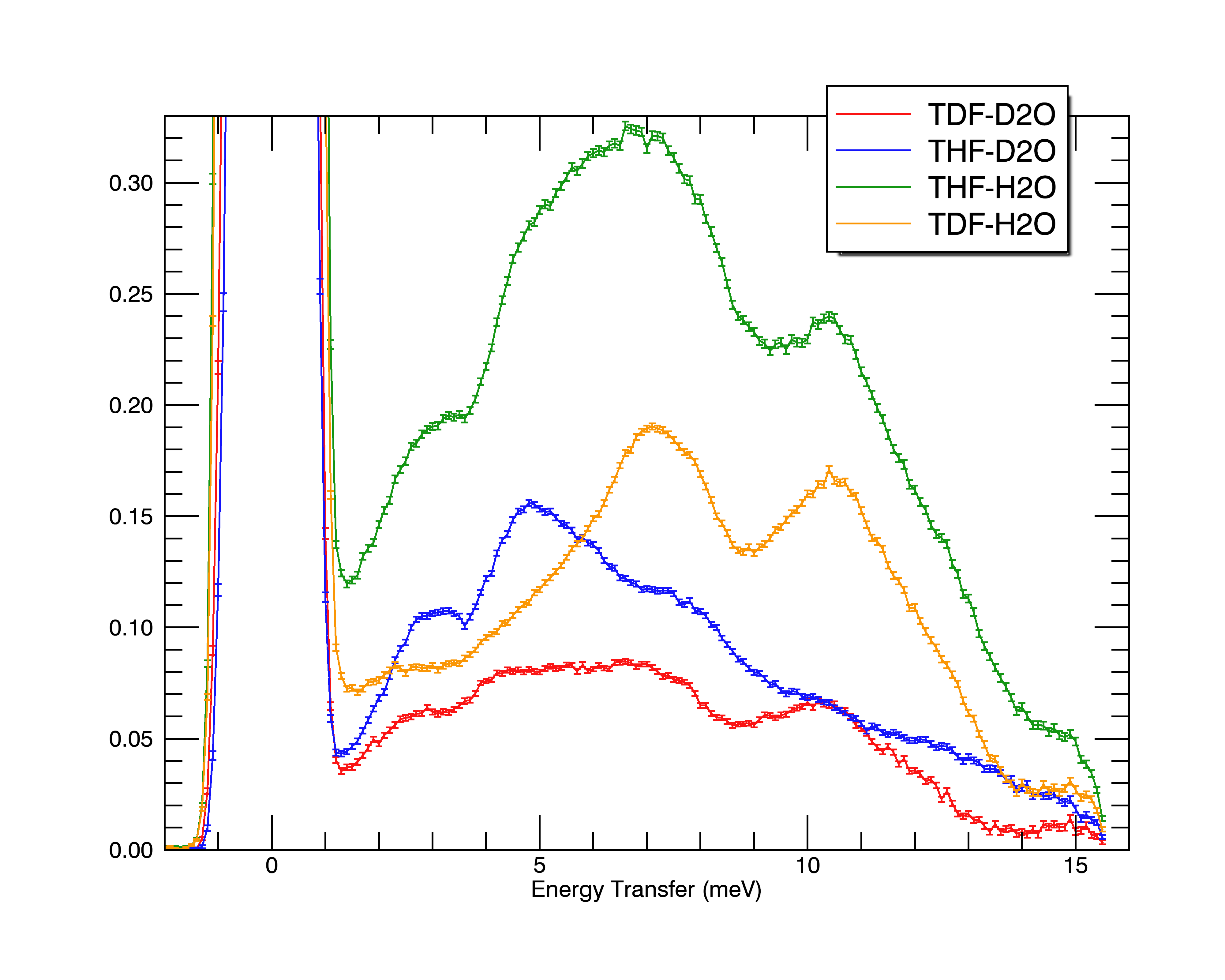} 
         \caption{Scattered intensities at certain energy transfers measured at IN5 for differently deuterated THF clathrate hydrates at an incident wavelength of 2\AA. }
         \label{fig:WP3_IN5}
     \end{subfigure}
     \begin{subfigure}[b]{0.43\textwidth}
         \centering
         \includegraphics[width=0.8\textwidth]{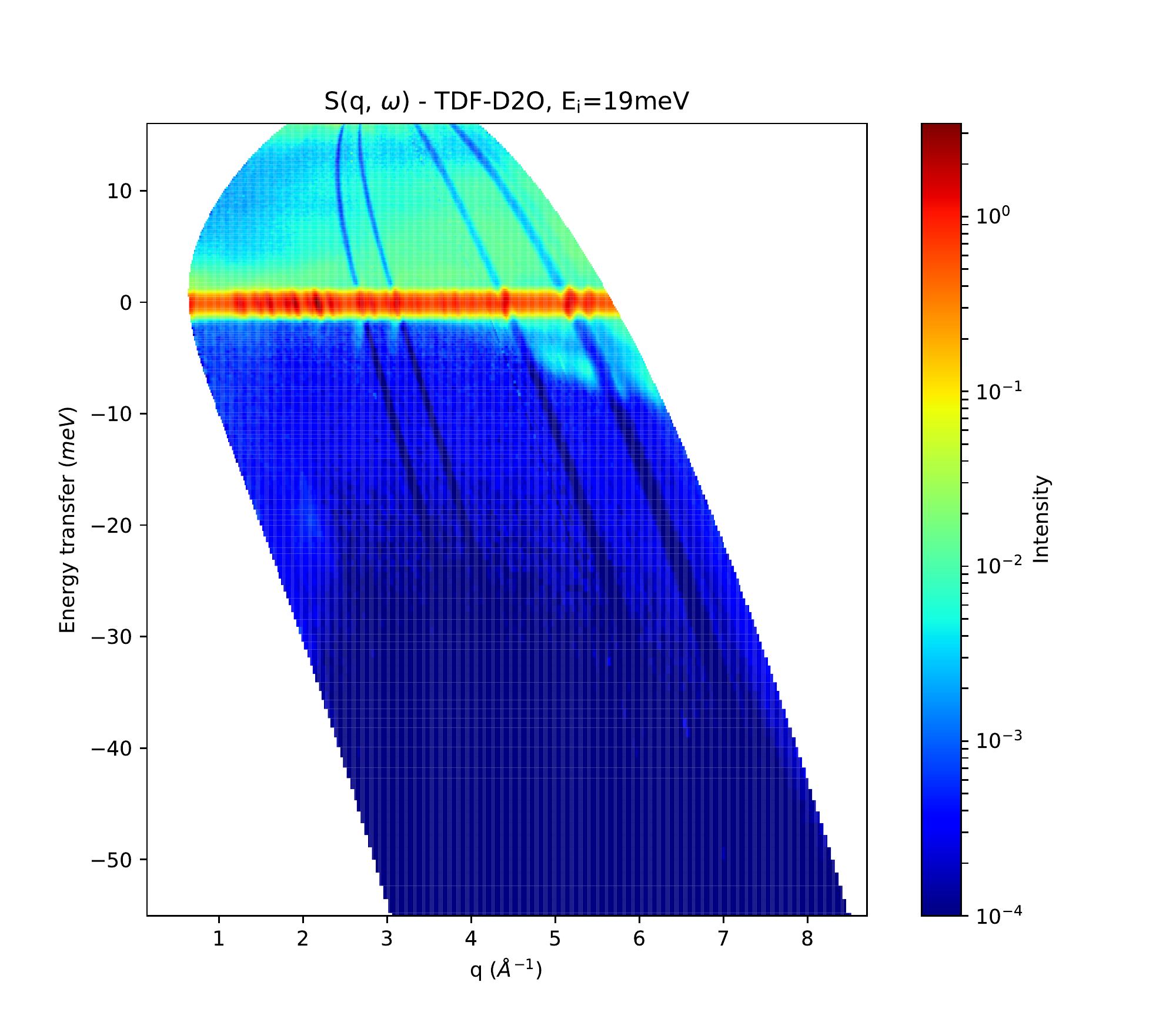} 
         \caption{Plot of $S(q,\omega)$ for TDF:D2O for a measurement at the instrument PANTHER for incoming neutron energy 19 meV. }
         \label{fig:WP3_Sqomega_PANTHER}
     \end{subfigure}
        \caption{}
        \label{fig:wp3_tof_results}
\end{figure}

In contrast to this experiment~\cite{conradInelasticScatteringSpectral2004}, that indicate the scattering functions $S(q,\omega)$ only in "arbitrary units", the experimental program of WP3 aims to determine $S(q,\omega)$ in absolutely calibrated units for various moderator materials.
Experiments have so far been performed on the inelastic neutron scattering time of flight (TOF) spectrometers IN5 and PANTHER at the ILL. A large range of neutron transfer energies and momenta could be covered. 
To absolutely determine $S(q,\omega)$ special attention was laid on the calibration measurements, which included measurements of empty sample cells and measurements of scattering intensities by vanadium standard samples of known mass and with the same shape as the samples of materials to be investigated. 
In Fig. \ref{fig:wp3_tof_results} some typical results of the performed experiments are shown. 

In addition, measurements have been performed at the high-intensity two-axis diffractometer D20.
An important result of these measurements has been that within the accuracy of the experiment the clathrate structure 
is completely developed from the stoichiometric mixture of its individual components and it does not depend on whether it is cooled down slowly or shock-freezed with liquid nitrogen.

Another crucial parameter for a possible novel moderator material is its transmissibility for VCNs, which corresponds to the mean free path in the medium.
These quantities were determined with a transmission experiment
of fully deuterated tetrahydrofuran (TDF) clathrate hydrate, using the VCN beam at the ILL's ultracold neutron facility PF2, where 
a wavelength range of 20 to 100~\AA~is provided.
Data analysis is ongoing and the results should provide the 
total cross-section of the material for the measured wavelength range, 
which is expected to be dominated by the incoherent scattering of the deuterium nuclei in the material.     
The main data set was acquired at a temperature of 5 K, but further transmission measurements have been carried out during the cooling and heating process to study the impact of excited phonon states with higher energy on the transmission behavior of the VCNs.

\section{Work Package 4: Moderators design}
WP4 has three main goals: to perform the neutronic design of a (i) cold, (ii) very cold, and (iii) ultracold source. 
The wavelength ranges for the three sources are roughly defined as 2-20~\AA~ for cold neutrons, 10-120~\AA~ for very cold neutrons, and \& >500 ~\AA~ for ultracold neutrons. 


The first year of the project was spent on two tasks: the definition of the figures of merit (FOM) for the design of the moderators, and an extensive set of simulations, using MCNP \cite{mcnp} for the neutronic design of the cold source, a high-intensity liquid deuterium moderator (LD$_2$).
For the first task, the definition of the FOM, it must be considered that the HighNESS sources are intended for two classes of applications,  neutron scattering instruments for condensed matter research and fundamental physics experiments, studied in WP7 and WP8, respectively.  The conventional range for neutron scattering experiments is between 2 and 20~\AA, but in HighNESS we extend this range to about 40~\AA. This is particularly interesting for SANS and spin-echo instruments, for which novel designs using longer wavelength neutrons are explored within the project. It was therefore decided to optimize the design of the moderator looking at the intensity from the emission surface, integrated in $\lambda$ between 2.5~\AA~and 40~\AA. Thus, the sources of interest for condensed matter research are the cold and very cold sources.
Concerning the FOM for fundamental physics, the main focus is on the NNBAR experiment, for which the sensitivity of the experiment is proportional to the number of neutrons, multiplied by the square of the time of flight. Hence, the FOM is given by the intensity of the cold neutrons, weighted by $\lambda^2$. For the UCN source, the FOM is given by maximizing the flux of cold neutrons that feed the UCN converter, e.g. 9~\AA~neutrons for a liquid He converter.
\begin{figure}[ht] 
  \centering
  \includegraphics[scale=0.11]{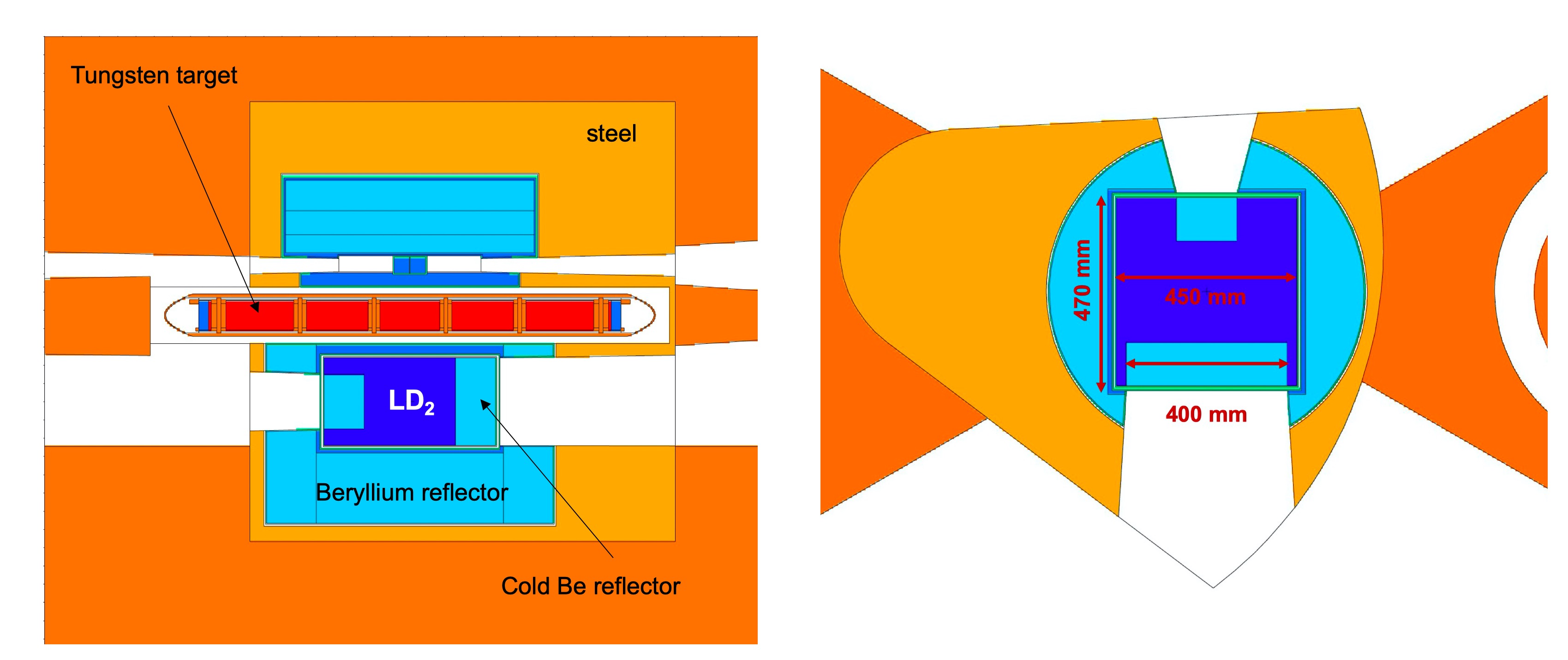}
  \caption{First geometry of the liquid deuterium moderator.}
  \label{fig:WP4Fig}
\end{figure}

The design of the high-intensity cold moderator (LD$_2$) should accommodate simultaneously instruments for the condensed matter science and fundamental physics, in fact most of the optimization process in the first year of the project was carried out to study a configuration where the NNBAR and neutron scattering instruments are placed on two opposite sides of the facility. In Fig.~\ref{fig:WP4Fig} we present the design obtained for which an engineering study is currently being carried out by WP5, as the first of several iterations between the neutronic and engineering teams. The moderator is box-shaped, with a large (24 cm high, 40 cm wide) opening on the NNBAR side, for maximum emission of neutrons, and a smaller (15 $\times$ 15 cm$^2$) opening on the other side, for neutron scattering. It was found that the presence of cold ($<$ 77~K) beryllium filter/reflectors placed on both sides increase the intensity of neutrons above 4~\AA. 
This first design delivers a large intensity of cold neutrons: the intensity of neutrons on the NNBAR side, integrated above 4 \AA, is about 7~10$^{15}$ n/s/sr, which roughly doubles the value previously calculated in Ref.~\cite{klinkby2014}, thanks in particular to the larger emission surface, and the presence of the cold filter/reflector. This high performance comes at the cost of a very large heat load on the cryogenic parts, of about 57 kW, for which the feasibility of the cooling is being studied by WP5.
In the second year of the project, it is planned to complete the design of the large LD$_2$ moderator, and start the design of the VCN and UCN sources.
 \label{ch:WP4}
\section{Work Package 5: Engineering}
\begin{figure}[h]
     \centering
     \begin{subfigure}[b]{0.42\textwidth}
         \centering
         \includegraphics[width=0.9\textwidth]{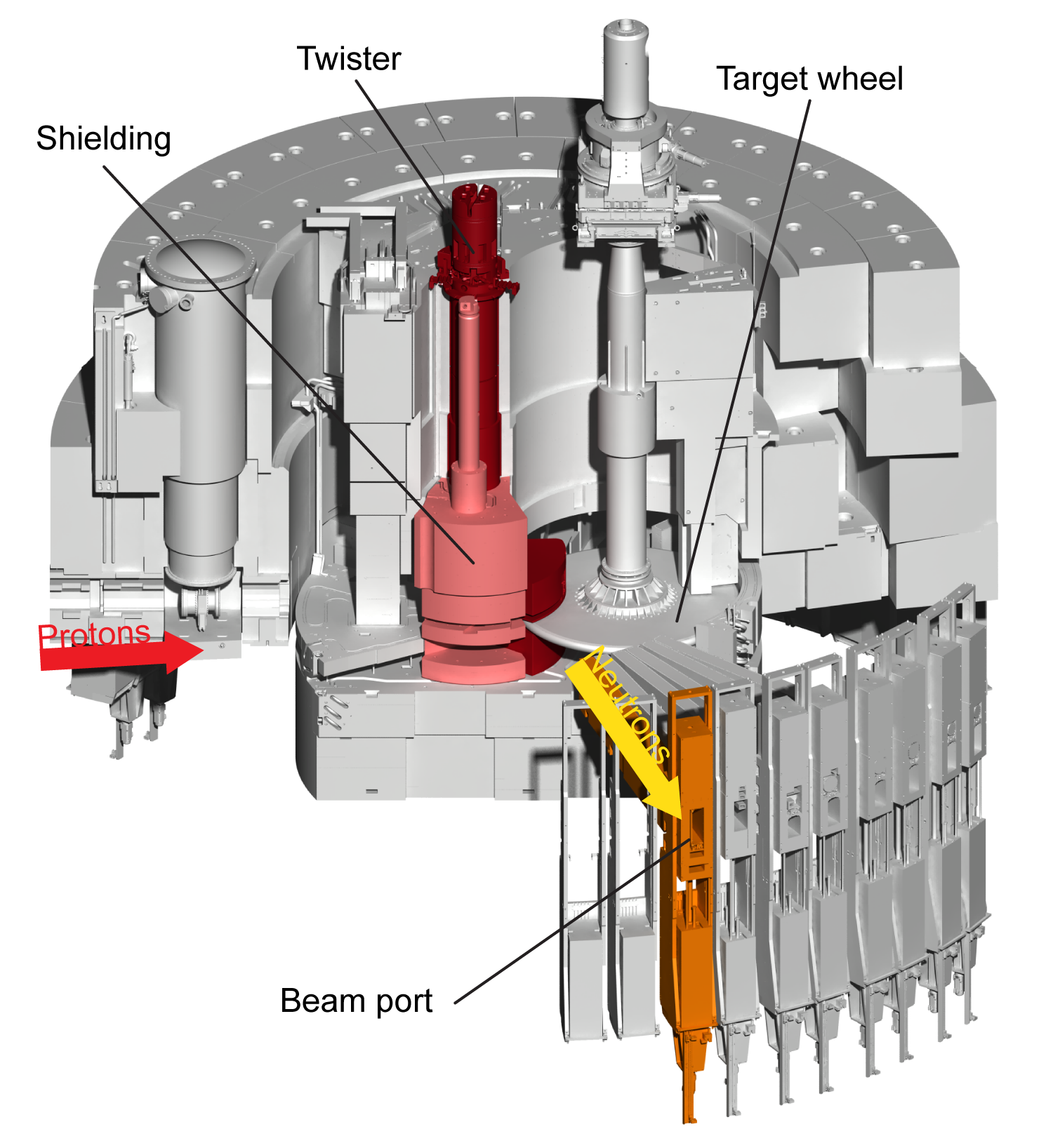} 
         \caption{ESS Target Monolith.}
         \label{fig:WP5Monolith}
     \end{subfigure}
     \begin{subfigure}[b]{0.42\textwidth}
         \centering
         \includegraphics[width=0.9\textwidth]{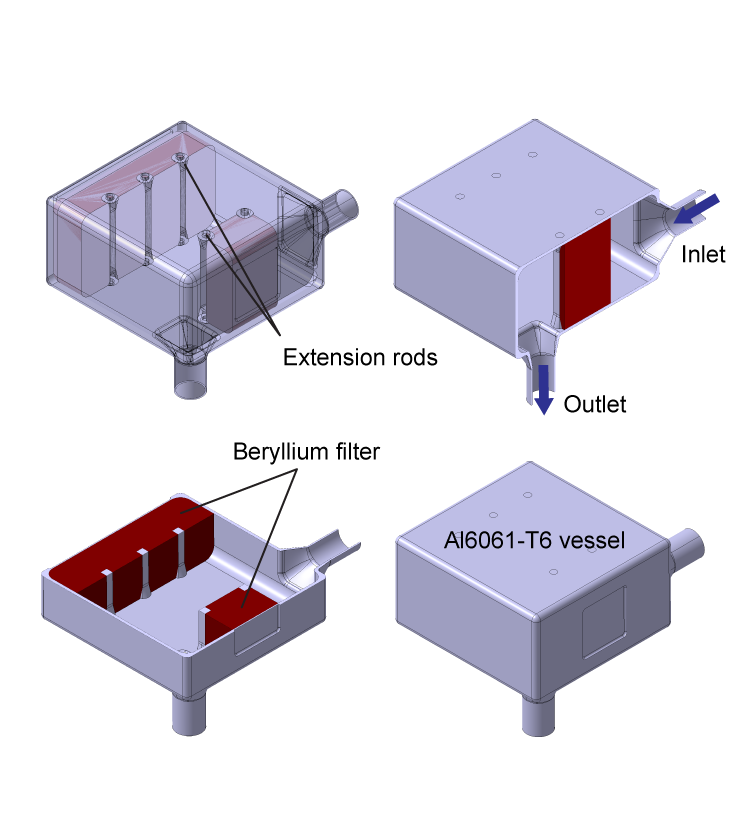} 
         \caption{Draft design of LD$_2$ moderator.}
         \label{fig:WP5LD2Mod}
     \end{subfigure}
        \caption{Preliminary engineering design of the lower moderator.}
        \label{fig:WP5Fig}
\end{figure}

The first generation of ESS moderators consists of two liquid parahydrogen low-dimensional moderators, located above the tungsten target wheel. The moderator support structure, the so-called Twister, allows using also the space below the target wheel for future moderator upgrades. Fig.~\ref{fig:WP5Monolith} shows the isometric view of the ESS target monolith. The colored components are subject to the planned upgrades of the ESS target station that is the subject of this paper.

The first sub-project in the framework of the HighNESS project is the engineering design of the volume LD$_2$ (see Fig.~\ref{fig:WP5LD2Mod}) described above, which will be installed below the target wheel in the Twister (see Fig.~\ref{fig:WP5Monolith}). The LD$_2$ will have a deuterium pressure of $p=5\,$bar and an average temperature of $T=21\,$K. The moderator vessel will be made of high-strength aluminum alloy Al6061-T6, which allows local stresses up to $S_{m}=87\,$MPa. The moderator will be surrounded by a vacuum jacket followed by a light water premoderator and a beryllium reflector. In addition, beryllium filter/reflectors are installed inside the cold moderator vessels at the neutron beam windows (see Fig. \ref{fig:WP5LD2Mod}). The biggest design challenge is to handle the enormous heat load into the LD$_2$ moderator of around ${\cal Q}=57\,$kW (see Sec.~\ref{ch:WP4}). A mass flow of at least $\dot{m}\geq 3400\,\frac{g}{s}$ liquid deuterium is needed to remove the particle heat and to stabilize the average temperature increase below $dT\leq 3\,$K. Assuming a flow velocity of up to $w\leq 5\,\frac{m}{s}$, a pipe diameter of d=70$\,$mm would be required, which have to fit into the existing Twister structure. The engineering design of the LD$_2$-moderator is still in progress and will be completed in 2022. In addition to the usual structural and fluid mechanics issues, the integrability must also be checked, as the components have to be integrated into an existing facility.\\
The second engineering sub-project in the framework of the HighNESS project will investigate the integrability of an ultracold neutron source into the existing high power facility. The preliminary location for the UCN was selected by the WP4 neutronics team and is part of the exchangeable shielding,  a so-called moderator cooling block (see Fig. \ref{fig:WP5Monolith}). This moderator cooling block is located outside of the main radiation field, whereby the UCN is exposed to a considerably smaller amount of particle heat compared to the LD$_2$ moderator. Finally, the main part of the last sub-project is to build an advanced reflector beam port prototype. This sub-project will start in 2022 and will be completed in 2023.
\section{Work Package 6: Advanced Reflector}
\begin{figure}[!h]
\centering
\includegraphics[width=0.38\columnwidth]{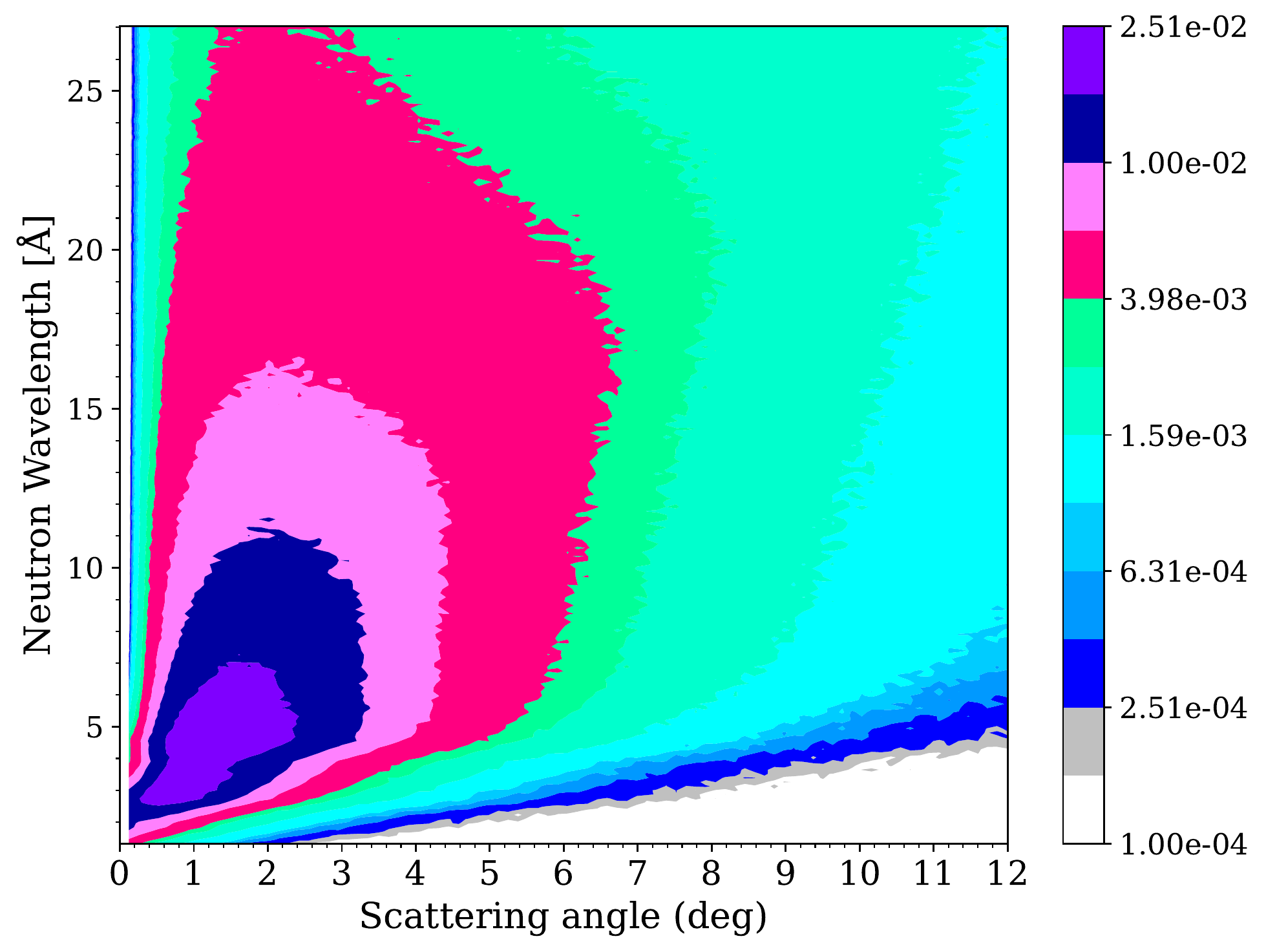}
\includegraphics[width=0.38\columnwidth]{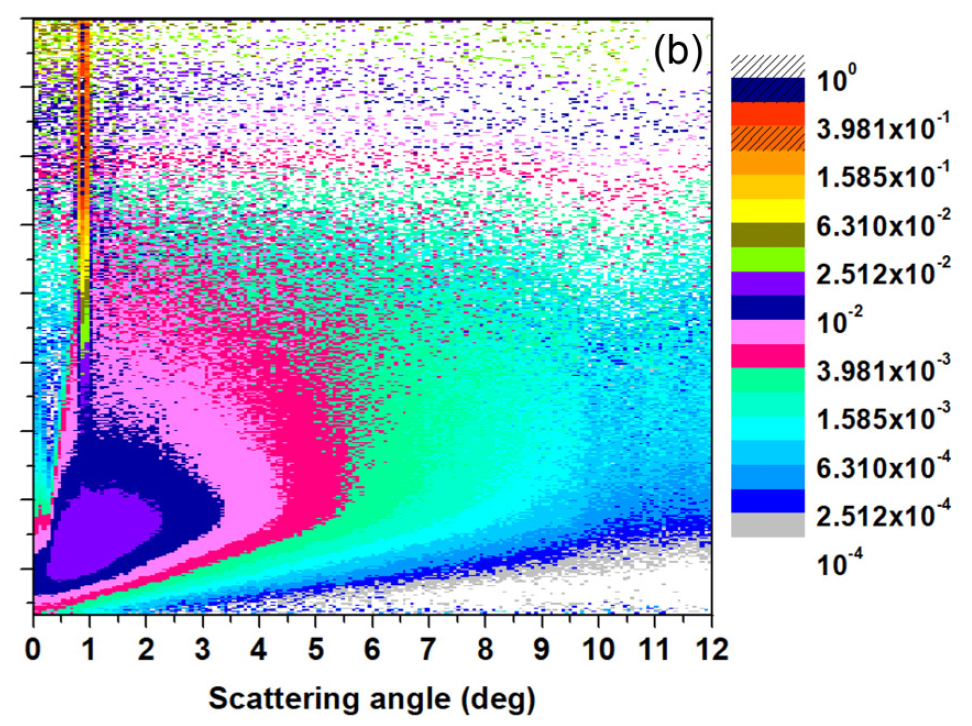}
\caption[Probability of neutron scattering from the surface of fluorinated nanodiamond powder as a function of the neutron wavelength and the scattering angle.]{Probability of neutron scattering from the surface of fluorinated nanodiamond powders as a function of the neutron wavelength and the scattering angle in the simulation with hard sphere model (left) and in \cite{nesvizhevsky_effect_2018} (right). The neutron incidence angle on the sample is $\mathbf 1 ^{\circ} $.}
\label{fig:2d_raw}
\end{figure}

The goal of WP6 is to study and optimize the design of advanced neutron reflectors to increase neutron delivery to the instruments. The focus is to reflect neutrons that intersect the beam extraction enclosure at higher incident angles with respect to the supermirrors critical angles. The knowledge gained in the WP2 and WP3 about the novel materials, then implemented in the relevant Monte Carlo software is essential for the success of this endeavor. In this sense, the development of a plugin for NCrystal\cite{ncrystal} to describe nanodiamond small-angle neutron scattering has been, in collaboration with WP2, a natural first step toward the optimal design. The reliability of the code has been extensively tested in different setups simulated in McStas\cite{willendrup2004mcstas} and the results compared with available experimental data. As an example, the reflectometry experiment described in \cite{nesvizhevsky_effect_2018} was reproduced in McStas, and results are shown in Fig.~\ref{fig:2d_raw}.

\begin{figure}[ht] 
  \centering
  \includegraphics[width=0.55\textwidth]{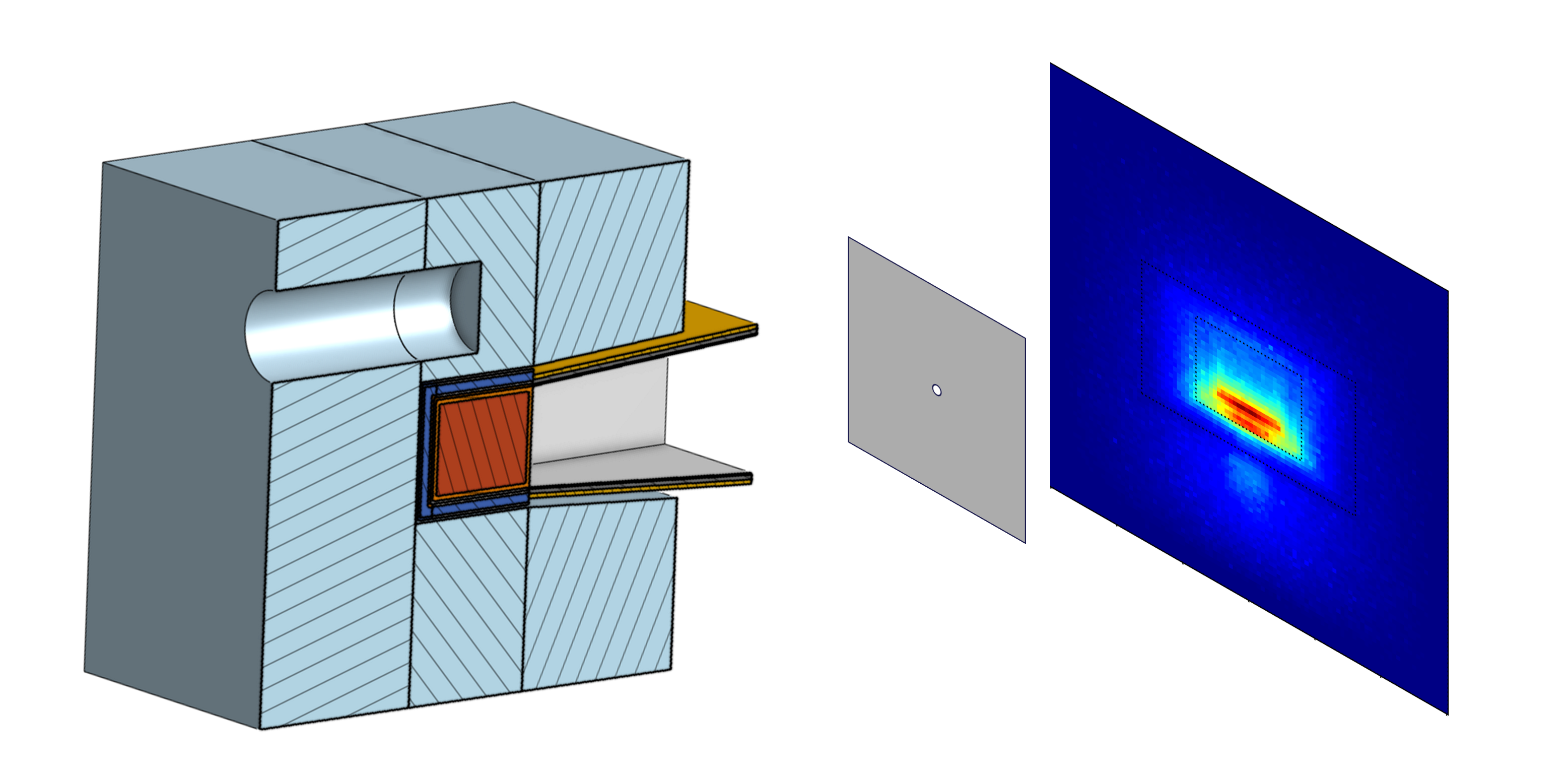}
  \caption{Proof of concept experiment simulated in MCNP. From left to right: moderator test facility model with advanced reflectors placed in the beam extraction system (grey surfaces), pinhole camera with an example of tally simulating the output of the detector.}
  \label{fig:WP6pinhole}
\end{figure}

Finally, a proof of concept experiment to test these advanced reflectors in a representative beam extraction configuration will be carried out at Budapest Neutron Center (BNC) where a Cold Moderator Test Facility (CMTF) is being built.
A dedicated experimental campaign at CMTF is under planning, to install an advanced neutron reflector in the vicinity of a cold moderator. This would allow validating neutron transport model calculations in a representative in-pile environment, hereby building confidence in modeling capabilities ahead of possible future implementation of advanced reflectors at ESS. A conceptual representation of the foreseen pinhole experiments at CMTF is shown in Fig.~\ref{fig:WP6pinhole}.
\section{Work Package 7: Condensed Matter Science}
The objective of WP7 entails the definition and optimization of the conceptual designs for neutron instruments for condensed matter research, that are expected to benefit from the future moderators designed in the HighNESS project. Some instrument types have been identified which benefit the most from the new source in the fields of small-angle neutron scattering (SANS), neutron imaging, and neutron spin-echo.
The expected increase in flux at longer wavelengths, offered by the second moderator, is the main motivation for the SANS instrument concept developed in the HighNESS project, aiming to complement the SANS instruments, LoKI \cite{andersen2020instrument} and SKADI \cite{andersen2020instrument, jaksch2021technical}, currently under construction at ESS. High flux in the  5-30~\AA~range will facilitate high resolution reaching into the low $Q$ regime probing long-ranging correlations e.g. in structural biology and soft matter (macromolecules, hierarchical structures) but also in hard matter research (e.g. flux line lattices). High flux in this key wavelength regime will further support investigations of small samples as well as rapid measurements. Rapid measurements with small beams will foster scanning-SANS investigations paving the way to probe local structures and structural variations throughout inhomogeneous samples. Faster measurements will on the other hand also support time-resolved SANS to study processes and structural transformations (e.g., reorientation dynamics of anisometric magnetic particles), that are not amenable today. On the other hand, the extended moderator surface area can additionally be utilized by attempting to design an instrument based on focussing optics.  Efficient collection of intensity from a large source area and focussing on a defined spot implies the additional promise of high flux and increased low $Q$ resolution worth being explored.  
To realize the above-envisioned goals, two alternative SANS instrument designs were conceptualized during the first year of the HighNESS project:

(1)	Conventional SANS geometry, with the use of a typical pin-hole collimator. The maximum collimation length is considered to be 30 m, with sample-to-detector distances of 3 m and 30 m and a total instrument length of 80 m. The long collimation offers high angular resolution also for large samples and is also complemented by high wavelength resolution due to the instrument length.

\begin{figure}[ht] 
  \centering
  \includegraphics[width=0.85\linewidth]{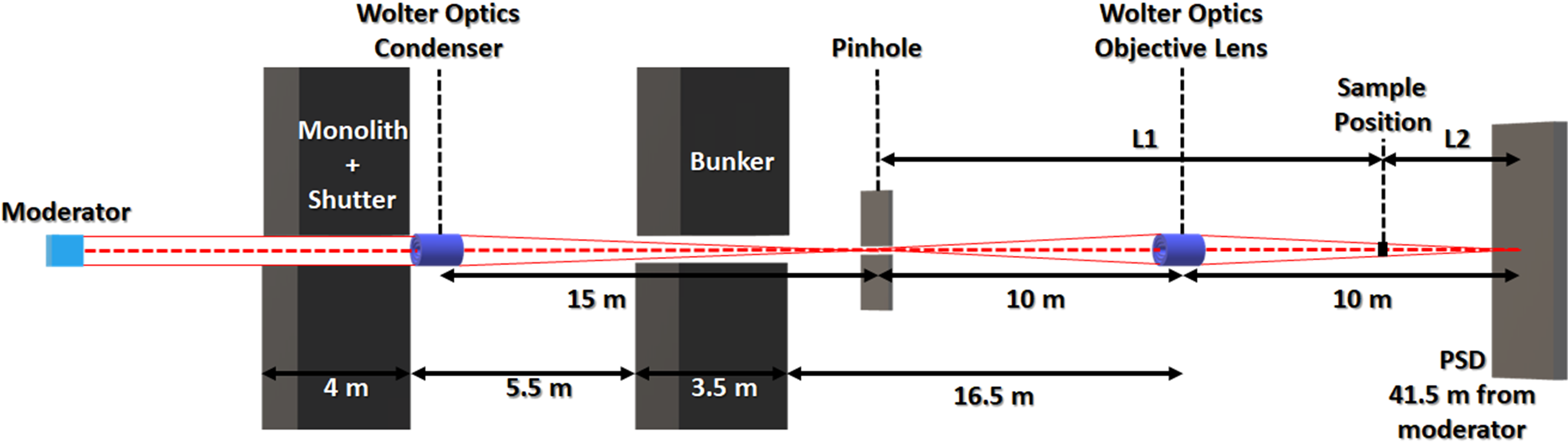}
  \caption{Wolter optics focusing SANS instrument layout (not in scale).}
  \label{fig:WP7Fig}
\end{figure}

(2)	Wolter optics~\cite{MILDNER2011S7}, focusing SANS instrument. The design consists of two Wolter mirror optics (see Fig. \ref{fig:WP7Fig}) to collect, transport, and focus the neutron beam. The first reflective, thus achromatic, optics is used as a condenser, collecting a large moderated neutron beam focusing it into a small well-defined spot. The second lens, the objective lens, is symmetric and refocuses the beam from the first focal spot onto the detector.

The conceptual design of a neutron imaging instrument at the new moderator is driven by complementing the currently constructed multi-purpose instrument ODIN \cite{andersen2020instrument, strobl2015scope}. The large moderator enables to build of a short instrument, nevertheless featuring a large beam cross section. The short distance implies that no guide optics are required which guarantees a homogeneous field of view. The short length and direct geometry lead to a high flux but very limited wavelength resolution. Thus, the instrument complements ODIN in providing superior flux and homogeneity most important for conventional attenuation contrast imaging of larger objects or highest resolution, spatially as well as temporally often utilizing the full provided wavelength spectrum. The high flux at long wavelength will additionally facilitate high performance measurements utilizing the long wavelength part of the spectrum only, enabling better quantification of materials with significant scattering, probing solely absorption. In addition, the moderate wavelength resolution available facilitates most efficient measurements e.g. of inelastic scattering contrast \cite{siegwart2019distinction}, depolarisation contrast \cite{strobl2019polarization} or small-angle scattering contrast, i.e. dark-field contrast \cite{strobl2008neutron}, where such moderate wavelength resolution absolutely suffices.\\
Work is also progressing on a number of conceptual ideas for an ultra-high resolution neutron spin-echo instrument, making full use of the increased flux of very long wavelength neutrons.\\
To optimize neutron guide systems of proposed instrument concepts for varying moderator parameters, the software tool {\it guide$\_$bot}  \cite{bertelsen2017automatic} is used, offering a systematic, fast and flexible solution for optimization. For the needs of the HighNESS project {\it guide$\_$bot}  \footnote{ The current development version of python guide$\_$bot can be found in the following link: https://pypi.org/project/guide-bot/.} is being modified and optimized, with respect to its former version \cite{bertelsen2017automatic}, into a python-based version, in particular, to also implement sophisticated Wolter optics optimizations.\\

\section{Work Package 8: Fundamental Physics}
\begin{figure}[ht]
  \centering
  \includegraphics[width=0.9\textwidth]{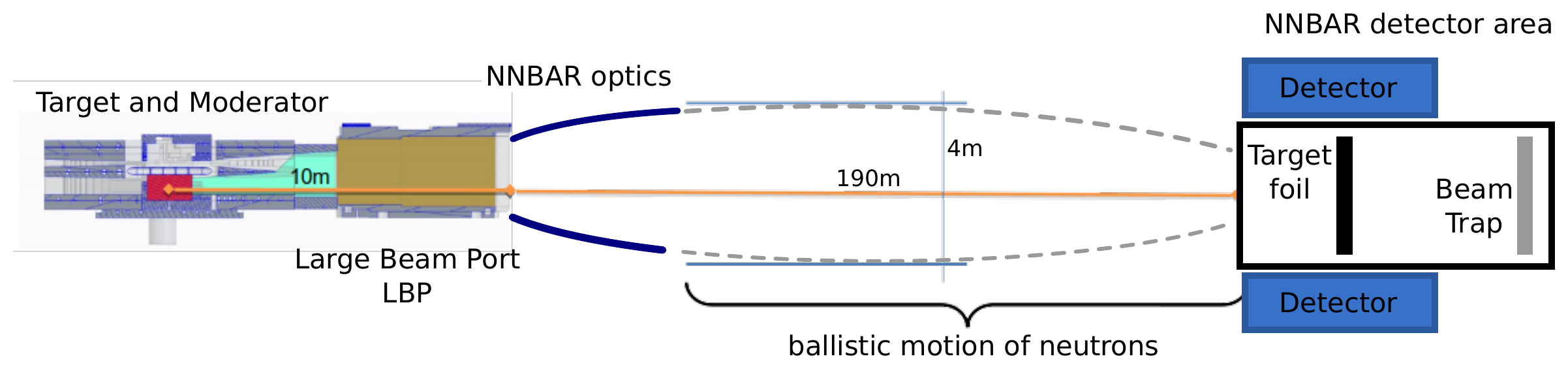}
  \caption{Schematic overview of the NNBAR experiment. The diagram is not to scale.} 
  \label{fig:wp8-over}
\end{figure}

WP8 is mostly dedicated to the design of the NNBAR experiment~\cite{Addazi:2020nlz}. This experiment would exploit the neutron flux available from the lower moderator to search for neutrons converting to anti-neutrons with a sensitivity that is three orders of magnitude greater than obtained in the last search with free neutrons~\cite{Baldo-Ceolin:1994hzw}. 
The experiment comprises neutron focusing and transport in a magnetically shielded region to a carbon target which is surrounded by a detector that can observe the annihilation between an anti-neutron and a target nucleon. This is schematically shown in Figure~\ref{fig:wp8-over}.

\begin{figure}[ht]
  \centering
  \includegraphics[width=0.3\textwidth]{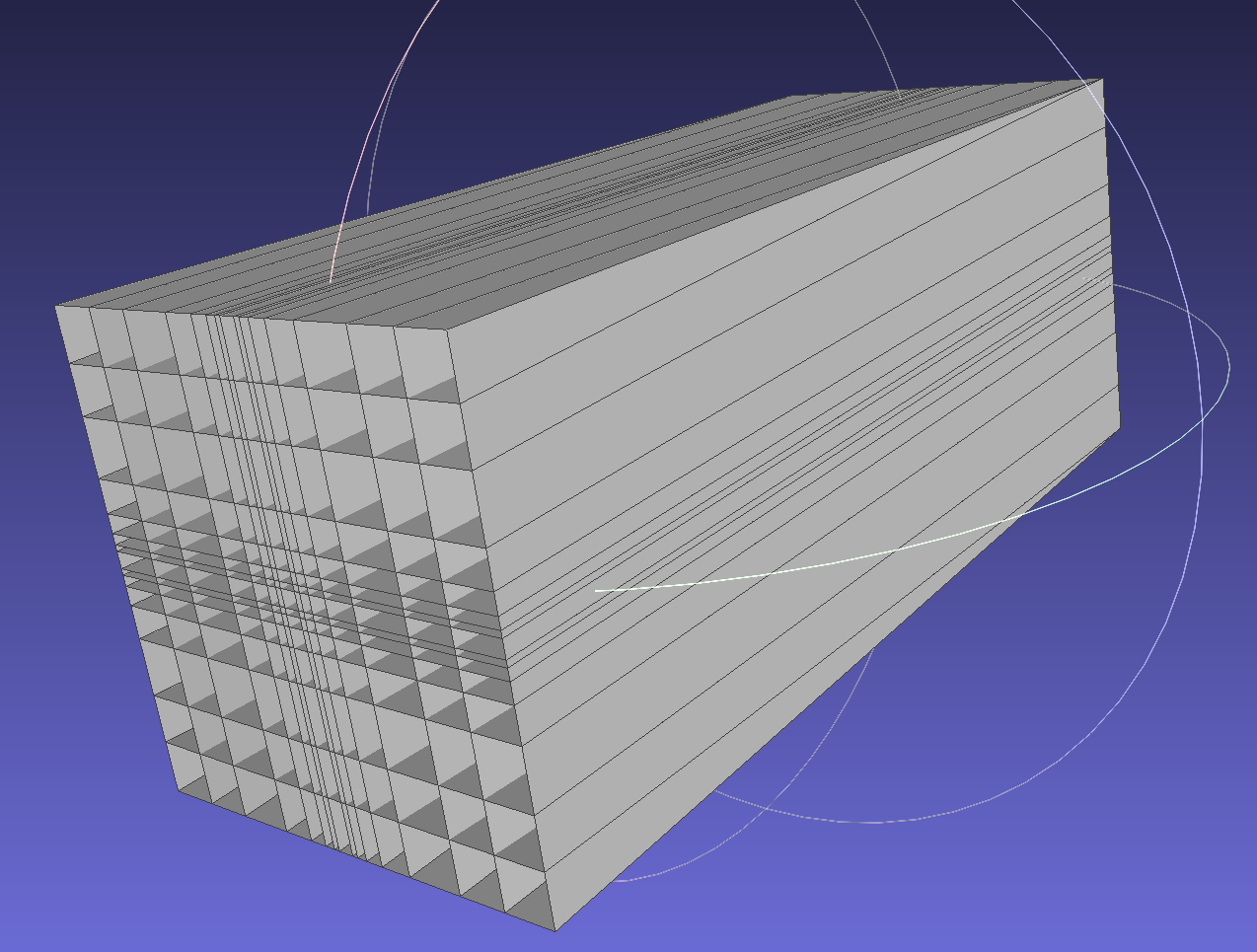}
  \includegraphics[width=0.3\textwidth]{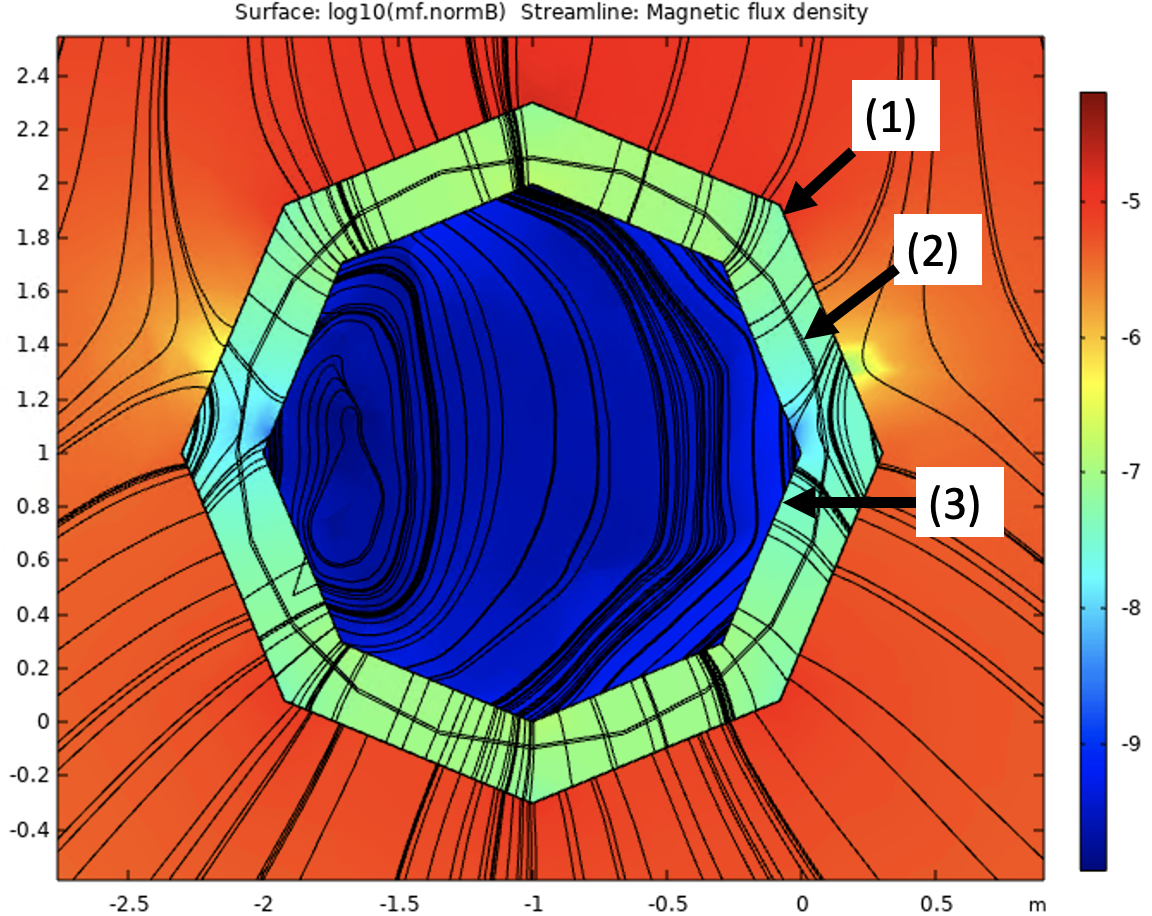}
   \includegraphics[width=0.4\textwidth]{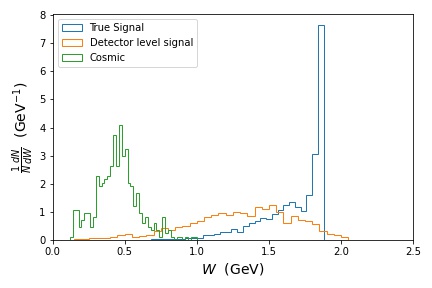}
  \includegraphics[width=0.4\textwidth]{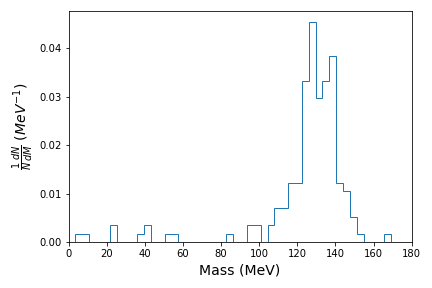}
  \caption{Top left: an example of a focusing double planar neutron optical component for the NNBAR experiment. Top right: Cut-through plot of a simulation of the DC field in the expected magnetic environment at the beam site. (1) is the outer shield, (2) the vacuum chamber and (3) the inner shield. Bottom left: the predicted invariant mass distribution for annihilation signal events and muon cosmic rays. The true and detector-smeared distributions are shown for the signal. 
  Bottom right: the detector-smeared diphoton mass distribution for neutral pion decays. }
  \label{fig:wp8-combined}
\end{figure}

A key aspect of the experiment is the neutron transport from the source to the detector and a major goal of WP8 is the design of an efficient system for this purpose. To evaluate various concepts and architectures of such systems  McStas ~\cite{willendrup2004mcstas} has been used to simulate different configurations.
A framework was developed that allows the creation of optical components in different geometries (i.e. elliptically nested or in Wolter type arrangement) from a few parameters corresponding to the geometrical constraints of the optics and its placement in the experiment. In Figure \ref{fig:wp8-combined} an example of such an optic is depicted.
The neutrons have to propagate in a region of low magnetic field (less than around 10~nT) to avoid breaking degeneracy between the neutron and the anti-neutron~\cite{Addazi:2020nlz}. A preliminary design of a magnetic shield has been simulated using a model from Ref.~\cite{Sun:lowfields}. This is based primarily on an outer and inner octagon-shaped passive shield of 1-2~mm thick sheets of mumetal, based on a concept demonstrated in Ref.~\cite{Wodey:fountainshield}. The mumetal shields together with means for degaussing~\cite{Altarev:equilibration} provide the required residual magnetic field of $<$~10~nT over the whole volume of interest. A cut of the field distribution for the expected conditions at the beamline for the octagonal shield with vacuum stainless steel vacuum chamber (one design option) is shown in Fig.~\ref{fig:wp8-combined}.


The annihilation detector would comprise a lead-glass electromagnetic calorimeter, supplemented with a hadronic range detector of scintillator staves. A time projection chamber provides high precision tracking and particle identification via measurements of the continuous energy loss $\frac{dE}{dx}$. The calorimeter and tracker are enclosed by an active cosmic muon shield made of scintillators and a passive shield overburden. Simulations of the detector using Geant4~\cite{GEANT4:2002zbu} have been developed within the NNBAR software framework~\cite{Barrow:2021deh}. Several discriminating variables have so far been studied. For example,  Fig.~\ref{fig:wp8-combined} shows the invariant mass for a signal annihilation event (at true and detector-smeared level) together with the prediction of a cosmic muon background event. The signal predictions were made with the model in Refs.~\cite{Golubeva:2018mrz,Barrow:2019viz} and cosmic ray background was simulated with the CRY~\cite{hagmann2007cosmic} program. Good discrimination is observed between signal and background processes. A key observable in suppressing backgrounds is the diphoton mass spectrum, also shown in Figure~\ref{fig:wp8-combined}, from which the presence of neutral pions can be inferred. The simulated detector shows good neutral pion reconstruction. The work program considers a range of background processes and observables to demonstrate the feasibility of a zero background experiment. 


\section{Work Package 9: Computing Infrastructure}
The objectives of WP9 are: to provide the required scientific computer infrastructure for the simulations needed in the HighNESS project and to make the tools developed available as cloud resources. Two simulation software packages need to be provided namely McStas and NCrystal  for thermal neutron transport. In 2021 the focus has been on providing computational resources for the participants in the HighNESS project providing the functionality required. This has taken the form of: (i) on-boarding users from the HighNESS community on the ESS Data Management and Software Center (DMSC) scientific compute cluster, (ii) providing storage capacity on the ESS scientific computer systems for HighNESS relevant data, (iii) keeping the McStas software up to date on the ESS DMSC scientific compute cluster (iv) enabling the possibility for running Jupyter notebooks as a batch job on the DMSC compute cluster. Aside from providing basic computational capacities for the HighNESS project, an effort has been undertaken to develop a proof of concept cloud solution for McStas simulation using JupyterHub. This has been achieved by utilizing a service put in place as part of the PaNOSC\footnote{Photon and Neutron Open Science Cloud \url{https://www.panosc.eu/}} project, namely the e-learning platform \url{pan-learning.org}. In \url{pan-learning.org} a training course can encompass the possibilities for performing limited size McStas simulations and a special course has been set up for HighNESS providing the possibility for HighNESS users to perform such simulations proving the viability of using JupyterHub as a framework for the cloud services to be provided. Such JupyterHubs must be provided as a more independent service than what is available in the pan-learning platform and needs to be connected to sufficient computational capacities which is currently not the case for pan-learning at moment.

\appendix

\section{Conclusions}
In summary, in the first year of the HighNESS project, several results have been achieved:  from the generation of scattering kernels for novel materials to the first moderator design and several different instruments concepts both for condensed matter science and for fundamental physics. Several developments are still ongoing and will be part of the Conceptual Design of the ESS upgrade which is the final objective of the HighNESS project.

\section{Acknowledgments}
This work was funded by the HighNESS project at the European Spallation Source. HighNESS is funded by the European Framework for Research and Innovation Horizon 2020, under grant agreement 951782.

\bibliographystyle{ans}
\bibliography{bibliography}
\end{document}